\documentclass[aps,showpacs,twocolumn]{revtex4}
\usepackage{epsfig}

\begin{document}
\title{Interpreting $Z_c(3900)$ and $Z_c(4025)/Z_c(4020)$ as charged tetraquark states}

\author{Chengrong Deng$^{a}{\footnote{crdeng@cqjtu.edu.cn}}$,
        Jialun Ping$^b{\footnote{jlping@njnu.edu.cn, corresponding author}}$,
        and Fan Wang$^c{\footnote{fgchenwang@nju.edu.cn}}$}

\affiliation{$^a$School of Mathematics and Physics, Chongqing Jiaotong University, Chongqing 400074, P.R. China}
\affiliation{$^b$Department of Physics, Nanjing Normal University, Nanjing 210097, P.R. China}
\affiliation{$^c$Department of Physics, Nanjing University, Nanjing 210093, P.R. China}

\begin{abstract}
In the framework of color flux-tube model with a four-body confinement potential, the lowest charged
tetraquark states $[Qq][\bar{Q}'\bar{q}']~(Q=c,b,q=u,d,s)$ are studied by using the variational method,
Gaussian expansion method. The results indicate that some compact resonance states can be formed, the
states can not decay into two color singlet mesons $Q\bar{q}'$ and $\bar{Q}'q$ through the breakdown
and recombination of color flux tubes but into $Q\bar{Q}'$ and $q\bar{q}'$. The four-body confinement
potential is an crucial dynamical mechanism for the formation of states, The decay mechanism is
similar to that of compound nucleus and therefore the states should be called ``color confined,
multi-quark resonance" states. The newly observed charged states $Z_c(3900)$ and $Z_c(4025)/Z_c(4020)$
can be accommodated in the color flux-tube model and can be interpreted as the $S$-wave tetraquark states $[cu][{\bar{c}\bar{d}}]$ with quantum numbers $I=1$ and $J=1$ and 2, respectively.
\end{abstract}

\pacs{14.20.Pt, 12.40.-y}

\maketitle

\section{Introduction}
In the past decade many charmonium and bottomonium (or charmonium-like and bottomonium-like)
states, denoted by $X$ and $Y$ particles, have been observed in experiments~\cite{review}.
Some states of them are not comfortably accommodated in quark model as $Q\bar{Q}'$ mesons
and therefore interpreted as exotic hadron states~\cite{review}, such as loose meson-meson
molecules, compact tetraquark states, hybrid quarkonia, and baryonia or hexaquark states
$q^3\bar{q}^3$. The situation has been further strengthened by the discovery of the charged
$Z$ particles~\cite{charge1,charge2}, because their minimum quark components must go beyond
conventional $Q\bar{Q}$ mesons and therefore are interpreted as exotic $Q\bar{Q}q\bar{q}'$
states. Very recently, the BES III Collaboration studied the process $e^+e^-\rightarrow\pi^+\pi^- J/\Psi$
at a center-of-mass energy of 4.26 GeV and reported a new charged charmonium-like structure
in the $\pi^{\pm}J/\Psi$ invariant spectrum, which is called $Z_c(3900)$ and has a mass
of $3899.0\pm3.6\pm4.9$ MeV and a width of $46\pm10\pm20$ MeV~\cite{z39001}. Almost at the same time,
the Belle observed a $Z(3895)^{\pm}$ state, with a mass of $3894.5\pm6.6\pm4.5$ MeV and a width of
$63\pm24\pm 26$ MeV in the $\pi^{\pm}J/\Psi$ invariant mass spectrum, in the process
$Y(4260)\rightarrow\pi^+\pi^- J/\Psi$ and ~\cite{z39002}. The state $Z_c(3900)$ has been
further confirmed by the CLEO-c Collaboration in the the decay $\psi(4160)\rightarrow\pi^+\pi^-J/\Psi$ with
a mass of $3886\pm4\pm2$ MeV and a width of $37\pm4\pm8$ MeV~\cite{z39003}. The states $Z_c(3900)$
and $Z(3895)^{\pm}$ have been observed by the BES III and  Belle Collaborations independently,
their masses and widths are agree well with each other within errors, which indicate $Z_c(3900)$
and $Z(3900)^{\pm}$ are the same state~\cite{zzc}. Subsquently, the BES III Collaboration studied
the process $e^+e^-\rightarrow\pi^{\pm} (D^*\bar{D}^*)^{\pm}$ at a center-of-mass energy of 4.26
GeV and reported a new charged charmonium-like structure, named as $Z_c^{\pm}(4025)$, with a mass
of $4026.3\pm2.6\pm3.7$ MeV and a width of $24.8\pm5.6\pm7.7$ MeV~\cite{z4025}. In addition, the
BES III Collaboration also observed an another charged state $Z_c(4020)$ very close to the
$(D^*\bar{D}^*)^{\pm}$ threshold with a mass of $4022.9\pm0.8\pm2.7$ MeV and a width of
$2.9\pm 2.7\pm 2.6$ MeV in the $\pi^{\pm}h_c$ invariant mass spectrum~\cite{z4020}.

Obviously, such charged states put forward a challenge for theoretical descriptions of
meson states a chance for tetraquark systems. A better understanding of the internal
structures of these and similar, yet unobserved, resonances may provide new insights
into the strong dynamics of multiquark systems, which is beneficial to understand
QCD low-energy behaviors. As a consequence, the great theoretical interests have been
aroused to comprehend the internal structures of those charged states with different
theoretical methods. So far, the theoretical interpretations can be classified into
three categories as follows. The first one is meson-meson molecules~\cite{molecule},
two mesons are separated at larger distances than the typical size of the mesons.
The interaction between two mesons can occur through exchange of mesons and gluons,
which is similar to nuclear force. Generally the interaction is weak and the mass of the
state is close to the threshold of two mesons. The second one is tetraquark states,
the four quarks may be divided into two clusters and form relatively tightly bound
diquark $[Qq]$ and antidiquark $[\bar{Q}'\bar{q}']$, which interact by the gluonic color
force and meson exchange force and decay through the rearrangement of the color
structure~\cite{tetraquark}. The last one is hadro-quarkonium, the heavy $Q\bar{Q}'$
pair forms a tightly bound system similar to the heavy quarkonium states, it is embedded
in a spatially large excited state of light mesonic matter and interacts with it by a
QCD analog of Van der Waals force~\cite{hadro-quarkonium,voloshin}. Which one is the
true picture of these charged states? More experimental and theoretical works are needed.

The present work aims at investigating the properties of the charged tetraquark states
with configuration $[Qq][\bar{Q}'\bar{q}']~(Q=c,b,~q=u,d,s)$ from the perspective
of a phenomenological model, a color flux-tube model, using the high-precision variational
method, Gaussian expansion method (GEM). In the model, the color confinement used is not the
sum of two-body interaction proportional to a color charge
$\mathbf{\lambda}^c_i\cdot\mathbf{\lambda}^c_j$ but a multibody one, which has been successfully
applied to study multiquark systems~\cite{y2175,cyclobutadiene,x1835,baryonia,scalar}.
The study attempts not only to describe the reported charged states and to enrich the list of the
possible charged states, but also to provide a new insight to charged states and to reveal the
underlying mechanism behind these novel phenomena. The calculation indicates that some
compact resonance states can be formed, in which the four-body confinement potential play a key
role, and the states can not decay into two color singlet mesons $Q\bar{q}'$ and $\bar{Q}'q$
through the strong interaction but into $Q\bar{Q}'$ and $q\bar{q}'$ through the breakdown
and recombination of color flux tubes. The newly observed charged states $Z_c(3900)$ and
$Z_c(4025)/Z_c(4020)$ can be interpreted as the S-wave tetraquark states $[cu][{\bar{c}\bar{d}}]$ or $[cd][{\bar{c}\bar{u}}]$ with quantum numbers $I=1$
and $J=1$ and 2, respectively, in the color flux-tube model.

The rest of the paper is organized as follows: the color flux-tube model and the corresponding
hamiltonian are given in Sec. II. Section III is devoted to the construction of the wave functions
of tetraquark states. The numerical results and discussions of the charged tetraquark states are
presented in Sec. IV. A brief summary is given in the last section.

\section{color flux-tube model and hamiltonian}

One important nature of quantum chromodynamics (QCD) is color confinement, whose understanding
continues to be a challenge in theoretical physics. Lattice QCD (LQCD) allows us to investigate the
confinement phenomenon in a nonperturbative framework and its calculations on $q\bar{q}$, $qqq$, and
tetraquark and pentaquark states reveal flux-tube or stringlike structures~\cite{latt1,latt2}. Such
flux-tube-like structures lead to a ``phenomenological'' understanding of color confinement,
the confinement potential is a multibody interaction which is proportional to the minimum of the
total length of flux tubes~\cite{latt1,latt2}.

The naive color flux-tube model has been developed based on the LQCD picture by taking into account a
multibody confinement potential with a harmonic interaction approximation; i.e., a sum of the square of
the length of flux tubes rather than a linear one is assumed to simplify the calculation~\cite{ping,ww}. The
approximation is justified because of the following two reasons: one is that the spatial variations
in separation of the quarks (lengths of the flux tube) in different hadrons do not differ significantly,
so the difference between the two functional forms is small and can be absorbed in the adjustable parameter,
the stiffness of a flux tube. The other is that we are using a nonrelativistic dynamics in the study.
As was shown long ago~\cite{goldman}, an interaction energy that varies linearly with separation between
fermions in a relativistic first order differential dynamics has a wide region in which a harmonic
approximation is valid for the second order (Feynman-Gell-Mann) reduction of the equations of motion.
The comparative studies also indicated that the difference between the quadratic confinement potential
and the linear one is very small~\cite{ping,ww}. The color flux-tube model can avoid the appearances of
the power law van der Waals forces between color-singlet hadrons and the anti-confinement in a color
symmetrical quark or antiquark pair in the traditional models, such as Isgur-Karl model and chiral quark
model~\cite{isgurkarl,chiral}, including two-body color-dependent confinement potential.

The color flux-tube structure of an ordinary hadron ($q\bar{q}$ meson or $q^3$ baryon) is unique and
trivial, the multibody quadratic confinement potential is equivalent to the sum of two-body
interactions in the color flux-tube model~\cite{baryonia}. In this sense, the color flux-tube model
is reduced to the traditional quark model and is therefore not a new model for ordinary hadrons.
For multiquark hadrons the situation is changed because the multiquark hadrons, if they really exist,
have various color flux-tube structures in the intermediate- and short-distance domains and the
corresponding multibody confinement potential can be not formulated into the sum of two-body ones.

The color flux-tube structures of multiquark hadrons are very important, because they own more low-energy
QCD information than ordinary hadrons, such as a quark pair with symmetric color representations.
The previous research on the light tetraquark spectrum indicated that a tetraquark system has at least
four possible color flux-tube structures: meson-meson molecule state $[q\bar{q}]_1[q\bar{q}]_1$,
hidden color octet-antioctet state $[[q\bar{q}]_8[q\bar{q}]_{\bar{8}}]_1$, diquark-antidiquark state
$[[qq]_3[\bar{q}\bar{q}]_{\bar{3}}]_1$ or $[[qq]_6[\bar{q}\bar{q}]_{\bar{6}}]_1$, and QCD cyclobutadiene
$[qq\bar{q}\bar{q}]_1$. The states with those different flux-tube structures are similarly called
QCD isomeric compounds, the details can be found in the Ref~\cite{cyclobutadiene}. Generally speaking,
a tetraquark state should be a mixture of all the possible structures. In order to avoid too
complicated numerical calculations, the diquark-antidiquark structure is considered in the present
work. Furthermore, the diquark-antidiquark structure is favored by many theoretical physicists.

Within the color flux-tube model, the confinement potential of the diquark-antidiquark structure
can be expressed as
\begin{eqnarray}
V^{C}(4) &=& K\left( (\mathbf{r}_{1}-\mathbf{y}_{12})^2
+(\mathbf{r}_2-\mathbf{y}_{12})^2+(\mathbf{r}_3-\mathbf{y}_{34})^2\right. \nonumber \\
&+&
\left.(\mathbf{r}_4-\mathbf{y}_{34})^2+\kappa_{d}(\mathbf{y}_{12}-\mathbf{y}_{34})^2\right),
\end{eqnarray}
where $\mathbf{r}_1$ and $\mathbf{r}_2$ represent two quarks' positions and $\mathbf{r}_3$ and $\mathbf{r}_4$ represent
two antiquarks' positions, the variational parameters $\mathbf{y}_{12}$ and $\mathbf{y}_{34}$ represent two junction
positions where three flux tubes meet. The parameter $K$ is the stiffness of a 3-dimension flux-tube, $\kappa_{d}K$ is
other compound color flux-tube stiffness, the relative stiffness parameter of the compound flux-tube $\kappa_{d}$~\cite{kappa}
\begin{equation}
\kappa_{d}=\frac{C_{d}}{C_3},
\end{equation}
where $C_{d}$ is the eigenvalue of the Casimir operator associated with the $SU(3)$ color representation
$d$ on either end of the color flux-tube, namely, $C_3=\frac{4}{3}$, $C_6=\frac{10}{3}$, and $C_8=3$.

For given quark (antiquark) positions $\mathbf{r}_i$, the junctions $\mathbf{y}_{12}$ and $\mathbf{y}_{34}$
can be obtained by minimizing the confinement potential. By introducing the following set of canonical
coordinates $\mathbf{R}_i$,
\begin{eqnarray}
\mathbf{R}_{1} & = &
\frac{1}{\sqrt{2}}(\mathbf{r}_1-\mathbf{r}_2),~
\mathbf{R}_{2} =  \frac{1}{\sqrt{2}}(\mathbf{r}_3-\mathbf{r}_4), \nonumber \\
\mathbf{R}_{3} & = &\frac{1}{ \sqrt{4}}(\mathbf{r}_1+\mathbf{r}_2
-\mathbf{r}_3-\mathbf{r}_4), \\
\mathbf{R}_{4} & = &\frac{1}{ \sqrt{4}}(\mathbf{r}_1+\mathbf{r}_2
+\mathbf{r}_3+\mathbf{r}_4). \nonumber
\end{eqnarray}
As a consequence, the minimum $V_{min}^c(4)$ of the confinement potential can be divided into three independence
harmonic oscillators and therefore have the following form,
\begin{eqnarray}
V^{C}_{min}(4) &=& K\left(\mathbf{R}_1^2+\mathbf{R}_2^2+
\frac{\kappa_{d}}{1+\kappa_{d}}\mathbf{R}_3^2\right)
\end{eqnarray}
Apparently, it is a four-body interaction and cannot be divided into the sum of six pairs two-body interactions.
For a two-body system, an ordinary meson, the confinement potential can be written as
\begin{eqnarray}
V^{C}_{min}(2) &=& K\mathbf{r}^2
\end{eqnarray}

The other prominent feature of QCD is the spontaneous breaking of the original $SU(3)_L\otimes SU(3)_R$ chiral
symmetry to $SU(3)_V$ at low momentum. In this region, up, down, and strange quarks obtain their constituent
quark masses and interact through Goldstone boson exchange (GBE). The chiral partner of pion, $\sigma$-meson,
is also introduced. In the heavy quark sector chiral symmetry is explicitly broken and therefore the GBE
interactions do not appear. The formulas of GBE potential $V_{ij}^B$ and $\sigma$-meson exchange potential $V_{ij}^{\sigma}$ are given by
\begin{eqnarray}
V_{ij}^B & = & V^{\pi}_{ij} \sum_{k=1}^3 \mathbf{F}_i^k
\mathbf{F}_j^k +V^{K}_{ij} \sum_{k=4}^7 \mathbf{F}_i^k
\mathbf{F}_j^k
\nonumber \\
& &+V^{\eta}_{ij} (\mathbf{F}^8_i \mathbf{F}^8_j\cos \theta_P
-\mathbf{F}^0_i \mathbf{F}^0_j\sin \theta_P)
\\
V^{\chi}_{ij} & = &
\frac{g^2_{ch}}{4\pi}\frac{m^3_{\chi}}{12m_im_j}
\frac{\Lambda^{2}_{\chi}}{\Lambda^{2}_{\chi}-m_{\chi}^2}
\mathbf{\sigma}_{i}\cdot
\mathbf{\sigma}_{j}\nonumber \\
& \times &\left( Y(m_\chi r_{ij})-
\frac{\Lambda^{3}_{\chi}}{m_{\chi}^3}Y(\Lambda_{\chi} r_{ij})
\right),~ \chi=\pi, K, \eta \nonumber\\
V^{\sigma}_{ij} & = &-\frac{g^2_{ch}}{4\pi}
\frac{\Lambda^{2}_{\sigma}}{\Lambda^{2}_{\sigma}-m_{\sigma}^2}
m_{\sigma}\left( Y(m_\sigma r_{ij})-
\frac{\Lambda_{\sigma}}{m_{\sigma}}Y(\Lambda_{\sigma}r_{ij})
\right)\nonumber
\end{eqnarray}
Where $Y(x)$ is standard Yukawa potential, $Y(x)=e^{-x}/x$. The symbols $\mathbf{F}$ and
$\mathbf{\sigma}$ are, respectively, the $SU(3)$ Gell-man and $SU(2)$ Pauli matrices. The $\chi$ in
$V^{\chi}_{ij}$ represents $\pi$, $K$ and $\eta$ mesons.

QCD perturbative effects are considered through introducing the one gluon-exchange (OGE) interaction.
OGE is responsible for the hyperfine splitting in the ordinary hadron mass spectrum, generally it takes
the following form
\begin{eqnarray}
V_{ij}^G & = & {\frac{1}{4}}\alpha_{s}\mathbf{\lambda}^c_{i}
\cdot\mathbf{\lambda}_{j}^c\left({\frac{1}{r_{ij}}}-
{\frac{2\pi\delta(\mathbf{r}_{ij})\mathbf{\sigma}_{i}\cdot
\mathbf{\sigma}_{j}}{3m_im_j}} \right),
\end{eqnarray}
where $\mathbf{\lambda}^c$ is the $SU(3)$ Gell-man matrices, and $m_i$ is the mass of $i$-th quark.
In order to obtain a unified description of light, strange and heavy mesons, a running strong coupling
constant has to be used. An effective scale-dependent strong coupling constant is used here~\cite{vijande},
\begin{equation}
\alpha_s(\mu_{ij})=\frac{\alpha_0}{\ln\left((\mu_{ij}^{2}+\mu_0^2)/\Lambda_0^2\right)}
\end{equation}
Where $\mu_{ij}$ is the reduced mass of two interactional quarks $q_i$ and $q_j$, namely
$\mu_{ij}=m_im_j/(m_i+m_j)$, $\Lambda_0$, $\alpha_0$ and $\mu_0$ are model parameters. The function
$\delta(\mathbf{r}_{ij})$ in OGE should be regularized, the regularization is justified based on the
finite size of the constituent quark and should be therefore flavor dependent~\cite{vijande,weistein},
\begin{equation}
\delta(\mathbf{r}_{ij})=\frac{1}{4\pi r_{ij}r_0^2(\mu_{ij})}e^{-r_{ij}/r_0(\mu_{ij})}
\end{equation}
The function $r_0(\mu_{ij})=\hat{r}_0/\mu_{ij}$, in which $\hat{r}_0$ is a model parameter determined by
ground state meson spectrum.

To sum up, the total hamiltonian $H_f$ can be expressed as the following form,
\begin{eqnarray}
H_f & = & \sum_{i=1}^f \left(m_i+\frac{\mathbf{p}_i^2}{2m_i}
\right)-T_{C}+\sum_{i>j}^f V_{ij}+V_{min}^C(f), \nonumber\\
V_{ij} & = & V_{ij}^B+V_{ij}^{\sigma}+V_{ij}^G
\end{eqnarray}
In the above expression of $H_f$, $f=2$ or $f=4$, $T_{c}$ is the center-of-mass kinetic energy, $\mathbf{p}_i$ is
the momentum of the $i$-th quark. The tensor forces and spin-orbit forces between quarks are omitted
in the model, because our primary interest is in the lowest energies and their contributions to the ground states
are small or zero. The present study involves up, down, strange, charm and bottom quarks, a different quark pair $q_iq_j$
($q_i\bar{q}_j$ or $\bar{q}_i\bar{q}_j$) therefore owns different interaction $V_{ij}$ listed in the following ,
\begin{eqnarray}
V_{ij}
 =\left\{
\begin{array}{ll}
\mbox{$V_{ij}^G$+$V_{ij}^{\pi}$+$V_{ij}^{\eta}$+$V_{ij}^{\sigma}$,~~~$q_iq_j=nn$}\\
\mbox{$V_{ij}^G$+$V_{ij}^{K}$+$V_{ij}^{\eta}$+$V_{ij}^{\sigma}$,~~$q_iq_j=ns$}\\
\mbox{$V_{ij}^G$+$V_{ij}^{\eta}$+$V_{ij}^{\sigma}$,~~~~~~~~~$q_iq_j=ss$}\\
\mbox{$V_{ij}^G$,~~~~~~~~~~~~~~~~~~~~~~$q_iq_j=Qn$}\\
\mbox{$V_{ij}^G$,~~~~~~~~~~~~~~~~~~~~~~$q_iq_j=Qs$}\\
\mbox{$V_{ij}^G$,~~~~~~~~~~~~~~~~~~~~~~$q_iq_j=QQ$}\\
\end{array}
\right. ,
\end{eqnarray}
where $n$ stands for the nonstrange light quarks, $u$ and $d$, $Q$ represents a charm or bottom quark.

\section{wave functions of charged tetraquark states}

In the diquark-antidiquark configuration, the wave function of a tetraquark state $[Qq][\bar{Q}'\bar{q}']$
can be written as a sum of the following direct products of color, isospin, spin and spatial terms,
\begin{eqnarray}
\Phi^{[Qq][\bar{Q}'\bar{q}']}_{IM_IJM_J} &=&
\sum_{\alpha}\xi_{\alpha}\left[ \left[
\left[\phi_{l_am_a}^G(\mathbf{r})\chi_{s_a}\right]^{[Qq]}_{j_a}
\left[\psi_{l_bm_b}^G(\mathbf{R})\right.\right.\right.\nonumber\\
& \times & \left.\left.\left.\chi_{s_b}\right]^{[\bar{Q}'\bar{q}']}_{j_b}
\right ]_{J_{ab}}^{[Qq][\bar{Q}'\bar{q}']}
F_{LM}(\mathbf{X})\right]^{[Qq][\bar{Q}'\bar{q}']}_{JM_J}\\
& \times & \left[\eta_{I_a}^{[Qq]}\eta_{I_b}^{[\bar{Q}'\bar{q}']}\right]_{IM_I}^{[Qq][\bar{Q}'\bar{q}']}
\left[\chi_{c_a}^{[Qq]}\chi_{c_b}^{[\bar{Q}'\bar{q}']}\right]_{CW_C}^{[Qq][\bar{Q}'\bar{q}']}, \nonumber
\end{eqnarray}
where $Q$ and $Q'$ stand for heavy quarks $c$ and $b$, $q$ and $q'$ stand for light quarks $u$, $d$ and $s$.
The codes of the quarks $Q$ and $q$, and the antiquarks $\bar{Q}'$ and $\bar{q}'$ are assumed to be 1 and 2,
and 3 and 4, respectively. In the center-of-mass frame, three relative motion coordinates, $\mathbf{r}$,
$\mathbf{R}$ and $\mathbf{X}$, can be expressed as
\begin{eqnarray}
\mathbf{r}&=&\mathbf{r}_1-\mathbf{r}_2,\ {}
\mathbf{R}=\mathbf{r}_3-\mathbf{r}_4,\nonumber\\
\mathbf{X}&=&\frac{m_1\mathbf{r}_1+m_2\mathbf{r}_2}{m_1+m_2}
-\frac{m_3\mathbf{r}_3+m_4\mathbf{r}_4}{m_3+m_4}%\\
\end{eqnarray}
The kinetic energy $T$ of the state $[Qq][\bar{Q}'\bar{q}']$ can therefore be written as
\begin{eqnarray}
T=\sum_{i=1}^4\frac{\mathbf{p}_i^2}{2m_i}-T_c=-\frac{\mathbf{p}^2_{\mathbf{r}}}{2\mu_{\mathbf{r}}}
-\frac{\mathbf{p}^2_\mathbf{R}}{2\mu_{\mathbf{R}}}
-\frac{\mathbf{p}^2_\mathbf{X}}{2\mu_{\mathbf{X}}},
\end{eqnarray}
where $\mu_{\mathbf{r}}$, $\mu_{\mathbf{R}}$, and $\mu_{\mathbf{X}}$ are the corresponding reduced masses.
$l_a$, $l_b$, and $L$ are the orbital angular momenta associated with the relative motions
$\mathbf{r}$, $\mathbf{R}$ and $\mathbf{X}$, respectively, and $J$ is total angular momentum.
$I_a,I_b$ are isospins of clusters $[Qq]$ and $[\bar{Q}'\bar{q}']$, respectively and $I$ is total isospin
of the state. $\alpha$ represents all possible intermediate quantum numbers,
$\alpha=\{l_k,s_k,j_k,J_{ab},L,I_k\}$, where $k=a,b$. $\chi_{s_k}$, $\eta_{I_k}$
and $\chi_{c_k}$ stand for spin, flavor and color wave functions of the diquark $[Qq]$ or the anti-diquark
$[\bar{Q}'\bar{q}']$, respectively. [~]'s denote Clebsh-Gordan coefficient coupling. The overall color singlet
can be constructed in two possible ways: $\chi_c^1=\bar{3}_{12}\otimes3_{34}$ and
$\chi_c^2=6_{12}\otimes\bar{6}_{34}$. The so-called ``good" diquark and the ``bad" diquark are both
included. The coefficient $\xi_{\alpha}$ is determined by diagonalizing the Hamiltonian.

The diquark $[Qq]$ (antidiquark $[\bar{Q}'\bar{q}']$) can be considered as a new compound object $\bar{\mathcal{Q}}$
($\mathcal{Q}'$) with no internal orbital excitations, and orbital excitations are assumed to occur only between
$\mathcal{Q}'$ and $\bar{\mathcal{Q}}$ in the present numerical calculations, which induces that such a tetraquark
state has a lower energy than the states with an internal orbital excitation. The orbital angular momenta
$l_a$ and $l_b$ are therefore assumed to be zero in the present work. Under this assumptions, $s_a=j_a$, $s_b=j_b$, $S=s_a+s_b=j_a+j_b=J_{ab}$, and $J=L+S$, where $S$ can be taken to be 0, 1, and 2. The parity of a tetraquark with
the diquark-antidiquark structure is $P=(-1)^L$.

To obtain a reliable numerical solution of a few-body problem, a high precision method is indispensable.
The GEM~\cite{gem}, which has been proven to be rather powerful in solving few-body problem, is used to
study four-body systems in the flux-tube model. In the GEM, three relative motion wave functions can be
expanded as,
\begin{eqnarray}
\phi^G_{l_am_a}(\mathbf{r}) &=& \sum_{n_a=1}^{n_{amax}}
c_{n_a}N_{n_al_a}r^{l_a}
e^{-\nu_{n_a}r^2}Y_{l_am_a}(\hat{\mathbf{r}}) \nonumber
\\
\psi^G_{l_bm_b}(\mathbf{R}) &=& \sum_{n_b=1}^{n_{bmax}}
c_{n_b}N_{n_bl_b}R^{l_b}
e^{-\nu_{n_b}R^2}Y_{l_bm_b}(\hat{\mathbf{R}})
\\
F^G_{LM}(\mathbf{X}) &=& \sum_{n_c=1}^{n_{cmax}}c_{n_c}
N_{n_cL}X^{L}e^{-\nu_{n_c}X^2}Y_{LM}(\hat{\mathbf{X}})\nonumber
\end{eqnarray}
Where $N_{n_al_a}$, $N_{n_bl_b}$ and $N_{n_{c}L}$ are normalization constants. Gaussian size parameters
are taken as the following geometric progression numbers:
\begin{eqnarray}
\nu_{n}=\frac{1}{r^2_n},& r_n=r_1x^{n-1},&
x=\left(\frac{r_{n_{max}}}{r_1}\right)^{\frac{1}{n_{max}-1}}.
\end{eqnarray}
The geometric progression leads to that $\nu_n$ is denser at intermediate- and short-range than at long-range,
so that it is suited to describe the dynamics mediated by intermediate- and short-range potentials.

\section{numerical results and discussions}

The mass spectrum of the ground states of mesons  from $\pi$ to $\Upsilon$ can be obtained by solving the
two-body Schr\"{o}dinger equation
\begin{eqnarray}
(H_2-E_{IJ})\Phi_{IJ}^{q\bar{q}}=0
\end{eqnarray}
with Rayleigh-Ritz variational principle in the color flux-tube model. The converged numerical results,
which are listed in Table I, can be reached by setting $r_1=0.3$ fm, $r_{n_{max}}=2.0$ fm and $n_{max}=5$.
The model parameters are fixed as follows. The masses $m_{\pi}$, $m_K$ and $m_{\eta}$ are taken their
experimental values, the mass $m_{\sigma}$ is determined through the PCAC relation $m^2_{\sigma}\sim
m^2_{\pi}+4m^2_{u,d}$~\cite{scadron}. The cutoff parameters are set to $\Lambda_{\pi}=
\Lambda_{\sigma}$=4.20 fm$^{-1}$ and $\Lambda_{K}=\Lambda_{\eta}$=5.20 fm$^{-1}$~\cite{vijande},
The mixing angle $\theta_{P}$ appears as a consequence of
considering the physical $\eta$ instead of the octet one~\cite{vijande}. The chiral coupling constant
$g_{ch}$ is determined from the $\pi NN$ coupling constant through
\begin{equation}
\frac{g_{ch}^2}{4\pi}=\left(\frac{3}{5}\right)^2\frac{g_{\pi NN}^2}{4\pi}\frac{m_{u,d}^2}{m_N^2}.
\end{equation}
The other parameters are fixed by fitting the mass spectrum of mesons. The fixed parameters and meson
mass spectrum are shown in Table I and II, respectively. Generally speaking, the color flux-tube model
can describe the meson spectrum well.
\begin{table}[ht]
\caption{The parameters in the color flux-tube model. $m_{\pi}$=0.7 fm$^{-1}$, $m_{K}$=2.51 fm$^{-1}$,
$m_{\eta}$=2.77 fm$^{-1}$, $m_{\sigma}$=2.92 fm$^{-1}$, $\Lambda_{\sigma}$=4.20 fm$^{-1}$,  $\Lambda_{K}=\Lambda_{\eta}$=5.20 fm$^{-1}$, $\theta_P=15^o$, $g_{ch}^2/{4\pi}$=0.43.}
\begin{tabular}{ccccccccc}\hline\hline
$m_{ud}$    & $m_{s}$ &  $m_{c}$  & $m_{b}$ &  $K$  & $\alpha_0$  &  $\Lambda_0$  & $\mu_0$  & $r_0$ \\
  MeV       &  MeV  &  MeV  &  MeV  &  MeV/fm$^{2}$ &          &  fm$^{-1}$   &  fm$^{-1}$   & MeV fm  \\
\hline
 280 &  510  &  1610   &  4980   & 149.3  &  5.60   &  0.016  &  0.147  & 0.153 \\
\hline\hline
\end{tabular}
\end{table}

\begin{table}[ht]
\caption{The ground state meson spectrum in the color flux-tube model, unit in MeV.}
\begin{tabular}{ccccccccccccc} \hline\hline
~Mesons~~        & ~~~Flavor~~~       &   ~~~$IJ^P$~~~       & ~Calculated~   &  ~Experimental~ \\
\hline
$\pi$            &   $n\bar{n}$       &   $10^-$              &      142       &     139         \\
$K$              &   $n\bar{s}$       &   $\frac{1}{2}0^-$    &      475       &     496         \\
$\rho$           &   $n\bar{n}$       &   $11^-$              &      785       &     775         \\
$\omega$         &   $n\bar{n}$       &   $01^-$              &      750       &     783         \\
$K^*$            &   $n\bar{s}$       &   $\frac{1}{2}1^-$    &      940       &     892         \\
$\phi$           &   $s\bar{s}$       &   $01^-$              &      1084      &     1020        \\
$D^{\pm}$        &   $c\bar{n}$       &   $\frac{1}{2}0^-$    &      1868      &     1869        \\
$D^*$            &   $c\bar{n}$       &   $\frac{1}{2}1^-$    &      1983      &     2007        \\
$D_s^{\pm}$      &   $c\bar{s}$       &   $00^-$              &      1963      &     1968        \\
$D_s^*$          &   $c\bar{s}$       &   $01^-$              &      2127      &     2112        \\
$\eta_c$         &   $c\bar{c}$       &   $00^-$              &      2890      &     2980        \\
$J/\Psi$         &   $c\bar{c}$       &   $01^-$              &      3103      &     3097        \\
$B^0$            &   $b\bar{n}$       &   $\frac{1}{2}0^-$    &      5288      &     5280        \\
$B^*$            &   $b\bar{n}$       &   $\frac{1}{2}1^-$    &      5322      &     5325        \\
$B_s^0$          &   $b\bar{s}$       &   $00^-$              &      5408      &     5366        \\
$B_s^*$          &   $b\bar{s}$       &   $01^-$              &      5456      &     5416        \\
$B_c$            &   $b\bar{c}$       &   $00^-$              &      6290      &     6277        \\
$B_c^*$          &   $b\bar{c}$       &   $01^-$              &      6393      &     ...         \\
$\eta_b$         &   $b\bar{b}$       &   $00^-$              &      9486      &     9391        \\
$\Upsilon(1S)$   &   $b\bar{b}$       &   $01^-$              &      9604      &     9460        \\
\hline\hline
\end{tabular}
\end{table}

The color flux-tube model with the model parameters listed in Table I is used to investigate the charge tetraquark
states $[Qq][\bar{Q}'\bar{q}']$. It should be emphasized that no any new parameter is introduced in the calculation
of the tetraquark states $[Qq][\bar{Q}'\bar{q}']$. The energies of the tetraquark states $[Qq][\bar{Q}'\bar{q}']$
can be obtained by solving the four-body Schr\"{o}dinger equation
\begin{eqnarray}
(H_4-E_{IJ})\Psi^{[Qq][\bar{Q}'\bar{q}']}_{IJ}=0.
\end{eqnarray}
The converged numerical results can be obtained by setting $n_{max}$=5, $N_{max}=5$ and $N^{\prime}_{max}=5$. The
minimum and maximum ranges of the bases are also 0.3 fm and 2.0 fm for coordinates $\mathbf{r}$, $\mathbf{R}$ and
$\mathbf{X}$, respectively.

In order to observe the underlying phenomenological features, the systematical calculations on the charged
tetraquark states $[Qq][\bar{Q}'\bar{q}']$ containing two heavy quarks and two light quarks are carried out.
We focus our attentions on the lowest charged states $[Qq][\bar{Q}'\bar{q}']$, and therefore the orbital angular momentum between two clusters is set to 0, then parity $P=+$, isospin $I=\frac{1}{2}$ or 1, and the total angular
momentum $J=0,1$ and 2. The lowest energies $E_{IJ}$ of the states $[Qq][\bar{Q}'\bar{q}']$ with quantum numbers
$IJ^P$ are given in Table III. The stability of these states can be identified by comparing the obtained
eigenvalues $E_{IJ}([Qq][\bar{Q}'\bar{q}'])$ with the corresponding meson-meson thresholds
$T_{M_1M_2}=M_1(Q\bar{q}')+M_2(\bar{Q}'q)$ and $T'_{M'_1M'_2}=M'_1(Q\bar{Q}')+M'_2(q\bar{q}')$ which are
calculated with the Hamiltonian $H_2$, namely $\Delta E=E_{IJ}([Qq][\bar{Q}'\bar{q}'])-T_{M_1M_2}$ and
$\Delta E'=E_{IJ}([Qq][\bar{Q}'\bar{q}'])-T'_{M'_1M'_2}$. If $\Delta E<0$ and $\Delta E'<0$ the states
$[Qq][\bar{Q}'\bar{q}']$ are bound states and cannot decay into two corresponding color singlet mesons
$Q\bar{q}'$, $\bar{Q}'q$ and $Q\bar{Q}'$, $q\bar{q}'$ under the strong interaction. While if
$\Delta E>0$ and $\Delta E'>0$, the states $[Qq][\bar{Q}'\bar{q}']$ may be resonances and can decay into
two corresponding color singlet mesons through the rupture and rearrangement of the color flux tubes in
the charged states $[Qq][\bar{Q}'\bar{q}']$.

It can be seen from Table III that the energies $E_{IJ}$ of all the states are higher than the threshold
of $M'_1(Q\bar{Q}')+M'_2(q\bar{q}')$, due to the large binding energies in the light mesons, $\pi$, $\rho$,
$K$  and $K^*$, which originating from the stronger interactions between two light quarks $q$ and $\bar{q}'$.
So the charged states $[Qq][\bar{Q}'\bar{q}']$ are hard to form bound tetraquark states and can always
decay into two mesons $Q\bar{Q}'$ and $q\bar{q}$, which is supported by the research~\cite{valcarce}.
On the contrary, the states $[QQ][\bar{q}\bar{q}]$ are easier to form stable tetraquark states due to that
they can only decay into two $Q\bar{q}$ mesons in the quark models~\cite{valcarce,ycyang}. From the results,
we also find that $\Delta E'$ of the low-spin ($S=0$ or 1) tetraquark states $[Qq][\bar{Q}'\bar{q}']$
are much higher, several hundreds MeVs, while $\Delta E'$ of the high-spin ($S=2$) states are several
tens of MeVs. Clearly, the differences come from the smaller masses of pseudo-scalar mesons.
%\begin{widetext}
%\begin{center}
\begin{table*}
\caption{The energies, unit in MeV, of the charge tetraquark states $[Qu][\bar{Q}'\bar{d}]$ or
$[Qu][\bar{Q}'\bar{d}]$ and $[Qu][\bar{Q}'\bar{s}]$ or $[Qd][\bar{Q}'\bar{s}]$ in the S-wave
$(L=0)$ between $[Qq]$ and $[\bar{Q}'\bar{q}']$, and the rms $\langle\mathbf{r}^2\rangle^{\frac{1}{2}}$,
$\langle\mathbf{R}^2\rangle^{\frac{1}{2}}$, and $\langle\mathbf{X}^2\rangle^{\frac{1}{2}}$ of the clusters
$[Qq]$, $[\bar{Q}\bar{q}]$, and $[Qq]$-$[\bar{Q}'\bar{q}']$, respectively, unit in fm.}
\begin{tabular}{c|cccc|cccc|cccc|cccc} \hline\hline
States         &   &$[cu][\bar{c}\bar{d}]$&$or$&$[cd][\bar{c}\bar{u}]$&&$[bu][\bar{b}\bar{d}]$&$or$&$[bd][\bar{b}\bar{u}]$&&$[cu][\bar{b}\bar{d}]$&$or$&$[cd][\bar{b}\bar{u}]$&&\\
$IJ^P$&$10^+$~&~$11^+$~&~$11^+$&~$12^+$~&~$10^+$~&~$11^+$~&~$11^+$~&$12^+$&~$10^+$~&~$11^+$~&~$11^+$~&~$12^+$\\

$E_{IJ}$       &   3778         &    3846        &    3846        &    3954          &     10330      &    10371          &    10371      &    10484
               &   7105         &    7144        &    7144        &    7255          \\
$T_{M_1M_2}$   &  $D\bar{D}$    & $D\bar{D}^*$   & $D^*\bar{D}$   &   $D^*\bar{D}^*$ &   $B\bar{B}$   &  $B\bar{B}^*$     &  $B^*\bar{B}$ &    $B^*\bar{B}^*$
               &  $D\bar{B}$    & $D\bar{B}^*$   &  $D^*\bar{B}$  &  $D^*\bar{B}^*$  \\
$\Delta E$     &   32           &    $-5$        &     $-5$       &    $-12$         &    $-246$      &     $-239$        &   $-239$      &    $-160$
               &   $-51$        &    $-46$       &     $-127$     &    $-50$         \\
$T'_{M'_1M'_2}$&  $\pi\eta_c$   &  $\pi J/\Psi$  &  $\rho\eta_c$  &  $\rho J/\Psi$   &  $\pi\eta_b$   & $\pi\Upsilon(1S)$ &  $\rho\eta_b$ & $\rho\Upsilon(1S)$
               & $\pi\bar{B}_c$ &$\pi\bar{B}_c^*$&$\rho\bar{B}_c$ &$\rho \bar{B}_c^*$\\
$\Delta E'$    &   746          &    601         &      171       &      66          &       702      &       625         &     100       &      95
               &   673          &    609         &      69        &      77          \\
$\langle\mathbf{r}^2\rangle^{\frac{1}{2}}$&~0.91&0.99&0.99&1.13&~0.91&0.92&0.92&1.08 &~0.92&0.96&0.96&1.11\\
$\langle\mathbf{R}^2\rangle^{\frac{1}{2}}$&~0.91&0.99&0.99&1.13&~0.91&0.92&0.92&1.08 &~0.91&0.94&0.94&1.11\\
$\langle\mathbf{X}^2\rangle^{\frac{1}{2}}$&~0.44&0.52&0.52&0.59&~0.24&0.26&0.26&0.28 &~0.37&0.41&0.41&0.45\\
\hline\hline
States         &  &$[cu][\bar{c}\bar{s}]$&$or$&$[cd][\bar{c}\bar{s}]$&&$[bu][\bar{b}\bar{s}]$&$or$&$[bd][\bar{b}\bar{s}]$
               &  &$[bu][\bar{c}\bar{s}]$&$or$&$[bd][\bar{c}\bar{s}]$&&$[cu][\bar{b}\bar{s}]$&$or$&$[cd][\bar{b}\bar{s}]$\\
$IJ^P$&~$\frac{1}{2}0^+$~&~$\frac{1}{2}1^+$~&~$\frac{1}{2}1^+$~&~$\frac{1}{2}2^+$~&~$\frac{1}{2}0^+$~&~$\frac{1}{2}1^+$~&~$\frac{1}{2}1^+$~&~$\frac{1}{2}2^+$~&
       ~$\frac{1}{2}0^+$~&~$\frac{1}{2}1^+$~&~$\frac{1}{2}1^+$~&~$\frac{1}{2}2^+$~&~$\frac{1}{2}0^+$~&~$\frac{1}{2}1^+$~&~$\frac{1}{2}1^+$~&~$\frac{1}{2}2^+$  \\
$E_{IJ}$       &   3990         &    4048        &    4048        &    4122          &    10545      &    10581       &    10581       &     10657
               &   7323         &    7353        &    7353        &    7429          &    7309       &    7345        &    7345        &     7418         \\
$T'_{M_1M_2}$  &  $D_s\bar{D}$  & $D_s\bar{D}^*$ & $D^*_s\bar{D}$ & $D^*_s\bar{D}^*$ & $B_s\bar{B}$  & $B_s\bar{B}^*$ & $B^*_s\bar{B}$ & $B^*_s\bar{B}^*$
               &  $\bar{D}B_s$  & $\bar{D}B^*_s$ & $\bar{D}^*B_s$ & $\bar{D}^*B^*_s$ & $D_s\bar{B}$  & $D_s\bar{B}^*$ & $D_s^*\bar{B}$ & $D_s^*\bar{B}^*$ \\
$\Delta E$     &    159         &    102         &     53         &    12            &    $-151$     &    $-149$      &     $-163$     &     $-121$
               &    47          &    29          &     $-38$      &   $-10$          &    58         &    60          &     $-70$      &     $-31$        \\
$T'_{M'_1M'_2}$&  $K\eta_c$     &  $K J/\Psi$    &   $K^*\eta_c$  &  $K^* J/\Psi$    &  $K\eta_b$    & $K\Upsilon(1S)$&  $K^*\eta_b$   & $K^*\Upsilon(1S)$
               &   $KB_c$       &   $KB^*_c$     &    $K^*B_c$    &   $K^*B^*_c$     & $K\bar{B}_c$  & $K\bar{B}_c^*$ & $K^*\bar{B}_c$ & $K^*\bar{B}_c^*$ \\
$\Delta E'$    &    625         &    470         &     218        &    79            &      584      &    502         &       155      &      113
               &    558         &    485         &     123        &    96            &      544      &    478         &       115      &      85          \\
$\langle\mathbf{r}^2\rangle^{\frac{1}{2}}$&~0.92&0.99&0.99&1.11&~0.91&0.93&0.93&1.06&~0.80&0.85&0.85&0.93&~0.93&0.97&0.97&1.06     \\
$\langle\mathbf{R}^2\rangle^{\frac{1}{2}}$&~0.80&0.87&0.87&0.96&~0.77&0.79&0.79&0.89&~0.91&0.94&0.94&1.07&~0.79&0.82&0.82&0.93     \\
$\langle\mathbf{X}^2\rangle^{\frac{1}{2}}$&~0.46&0.53&0.53&0.58&~0.25&0.27&0.27&0.28&~0.39&0.42&0.42&0.46&~0.38&0.42&0.42&0.45     \\
\hline\hline
\end{tabular}
\end{table*}
%\end{center}
%\end{widetext}

The energies $E_{IJ}$ of the states $[Qq][\bar{Q}'\bar{q}']$ are also compared with the threshold
$T_{M_1M_2}=M_1(Q\bar{q}')+M_2(\bar{Q}'q)$ in Table III, it can be found that the energies of many
states lies below the thresholds $T_{M_1M_2}$, $\Delta E<0$. Let us
first pay our attentions to the charged states $[cu][\bar{c}\bar{d}]$ or $[cd][\bar{c}\bar{u}]$
due to the observations of the charged states $Z_c(3900)$ and $Z_c(4025)$/$Z_c(4020)$ in
experiments~\cite{z39001,z39002,z39003,zzc,z4025,z4020}, the isospin of the charged states
$[cu][\bar{c}\bar{d}]$ or $[cd][\bar{c}\bar{u}]$ must be $I=1$. In the present calculation,
The energies of the states with $J=1$ and $2$ are lower than the threshold
$T_{M_1M_2}=M_1(Q\bar{q}')+M_2(\bar{Q}'q)$ by 5 MeV and 12 MeV, respectively. So
these two states cannot decay into $D^*\bar{D}$ or $D\bar{D}^*$ and $D^*\bar{D}^*$ through strong
interactions in the color flux-tube model. After taking the meson mass differences between the
calculated data and the experimental data into account, see Table I, the energies $E_{IJ}$ with
$J=1$ and $2$ should be $3871$ MeV and $4002$ MeV in the color flux-tube model, respectively.
Therefore the energies of the charged states $[cu][\bar{c}\bar{d}]$ with $J=1$ and $2$ are in
good agree with the experimental data of the charged sates $Z_c(3900)$ and
$Z_c(4025)/Z_c(4020)$~\cite{z39001,z39002,z39003,zzc,z4025,z4020}. It is possible that the
dominant component of the charged sates $Z_c(3900)$ and $Z_c(4025)/Z_c(4020)$ may be the hidden color
tetraquark states $[cu][\bar{c}\bar{d}]$ with $I=1$ and $J=1$ and $2$, respectively. Our points of view
on interpreting the charged states as tetraquark states are supported by the other
research~\cite{tetraquark}. With regarding to the state $[cu][\bar{c}\bar{d}]$ with $IJ=10$,
the energy is 32 MeV higher than the $D\bar{D}$ threshold in our calculation. Considering the
meson mass differences between the calculated data and the experimental data, the predicted energy
is 3780 MeV, which is very close to the result, 3785 MeV in the hadro-quarkonium picture~\cite{voloshin},
where the state $[cu][\bar{c}\bar{d}]$ with quantum numbers $I^GJ^{P}=1^-0^+$ is called $W_c$.

The charge tetraquark states consisting of both hidden charm and open strange components,
namely the state $[cu][\bar{c}\bar{s}]$ or $[cs][\bar{c}\bar{u}]$, are also investigated
in the color flux-tube model. One can find from Table III that the energies of the states
with $J=0$ and $J=1$ are much higher than the thresholds $D\bar{D}_s$ and $D^*\bar{D}_s$,
respectively. While the energies of the states $[cu][\bar{c}\bar{s}]$ with $J=1$ and $J=2$
are close to the thresholds $D\bar{D}^*_s$ and $D^*\bar{D}^*_s$, respectively. Considering
the meson mass differences between the calculated data and the experimental data,
the predicted $[cu][\bar{c}\bar{s}]$ with $J=1$ and $J=2$ possess energies are, respectively,
4033 MeV and 4131 MeV in our model. The initial single chiral particle emission mechanism
was used to study the charged charmoniumlike structures with hidden-charm and open-strange
and obtained that the masses of those structures are also near the thresholds of
$D\bar{D}^*_s/D^*\bar{D}_s$ and $D^*\bar{D}^*_s/D_s^*\bar{D}^*$~\cite{dychen}. The future
experiments are suggested to carry out the search for these charged charmoniumlike structures
with hidden-charm and open-strange channels.

It can be seen from Table III that the bigger the mass ratios ${M_Q}/{m_q}$ and
${M_{\bar{Q}'}}/{m_{\bar{q}'}}$ in the different charged states $[Qq][\bar{Q}'\bar{q}']$ but
with the same $IJ^P$, the smaller the corresponding binding energies $\Delta E$, while $\Delta E's$
do not change much. The reason of the tendency is that the big mass of heavy
quarks depresses the motion domain (decrease the contribution from the color confinement) and
reduces the kinetic energy, while other interactions have not great changes. The charged states
$[bu][\bar{b}\bar{d}]$ or $[bd][\bar{b}\bar{u}]$ and $[bu][\bar{b}\bar{s}]$
or $[bs][\bar{b}\bar{u}]$ have rather strong binding. Their masses are too small comparing with
experimental data of the charged states $Z_b(10610)$ and $Z_b(10650)$~\cite{charge2}.
For the other charged states $[cu][\bar{b}\bar{d}]$ or $[cd][\bar{b}\bar{u}]$, $[bu][\bar{c}\bar{s}]$
or $[bd][\bar{c}\bar{s}]$ and $[cu][\bar{b}\bar{s}]$ or $[cd][\bar{b}\bar{s}]$, the energies are very
close to the corresponding thresholds.
\begin{center}
\begin{table}
\caption{The distances $\langle\mathbf{r}_{ij}^2\rangle^{\frac{1}{2}}$ between the $i$-th particle and the $j$-th
particle in the states $[cu][\bar{c}\bar{d}]$ or $[cd][\bar{c}\bar{u}]$, where $\mathbf{r}_{ij}=\mathbf{r}_{i}-\mathbf{r}_{j}$, unit in fm.}
\begin{tabular}{ccccccccccccccccc} \hline\hline
~Distances&~$\langle\mathbf{r}_{12}^2\rangle^{\frac{1}{2}}$~&~$\langle\mathbf{r}_{34}^2\rangle^{\frac{1}{2}}$~&~$\langle\mathbf{r}_{24}^2\rangle^{\frac{1}{2}}$~
          &~$\langle\mathbf{r}_{13}^2\rangle^{\frac{1}{2}}$~&~$\langle\mathbf{r}_{14}^2\rangle^{\frac{1}{2}}$~&~$\langle\mathbf{r}_{23}^2\rangle^{\frac{1}{2}}$~\\
$IJ=10$   &   0.91         &    0.91        &    1.19        &    0.48          &     0.90      &    0.90   \\
$IJ=11$   &   0.99         &    0.99        &    1.30        &    0.56          &     1.00      &    1.00       \\
$IJ=12$   &   1.13         &    1.13        &    1.48        &    0.63          &     1.14      &    1.14   \\
\hline\hline
\end{tabular}
\end{table}
\end{center}

%the Belle and BABAR and forthcoming BelleII and
%SuperB Collaborations will be good platforms to carry out the search for these charged structures.

In order to obtain the spacial configurations of the charged states, the rms $\langle\mathbf{r}^2\rangle^{\frac{1}{2}}$, $\langle\mathbf{R}^2\rangle^{\frac{1}{2}}$, and $\langle\mathbf{X}^2\rangle^{\frac{1}{2}}$, which stand for the sizes of the clusters $[Qq]$,
$[\bar{Q}'\bar{q}']$ and the distance between the clusters $[Qq]$ and $[\bar{Q}'\bar{q}']$, are
calculated by using the obtained eigen wavefunctions and also shown in Table III. It can be seen
that the higher the spins of the clusters $[Qq]$ and $[\bar{Q}'\bar{q}']$, the bigger the rms
$\langle\mathbf{r}^2\rangle^{\frac{1}{2}}$, $\langle\mathbf{R}^2\rangle^{\frac{1}{2}}$,
and $\langle\mathbf{X}^2\rangle^{\frac{1}{2}}$. The bigger the masses of the clusters $[Qq]$ and 
$[\bar{Q}'\bar{q}']$, the smaller the $\langle\mathbf{r}^2\rangle^{\frac{1}{2}}$, 
$\langle\mathbf{R}^2\rangle^{\frac{1}{2}}$, and $\langle\mathbf{X}^2\rangle^{\frac{1}{2}}$.
The $\langle\mathbf{r}^2\rangle^{\frac{1}{2}}$ and $\langle\mathbf{R}^2\rangle^{\frac{1}{2}}$ are 
around 1 fm, while the $\langle\mathbf{X}^2\rangle^{\frac{1}{2}}$ is much smaller than the 
$\langle\mathbf{r}^2\rangle^{\frac{1}{2}}$ and $\langle\mathbf{R}^2\rangle^{\frac{1}{2}}$. 
In this way, the charged states $[Qq][\bar{Q}'\bar{q}']$ should be compact tetraquark
states because the two clusters $[Qq]$ and $[\bar{Q}'\bar{q}']$ have a large overlap. 
To make the spatial structure of tetraquark state more clear, the distances between
any two particles in the states $[cu][\bar{c}\bar{d}]$ are calculated and listed in 
Table IV, in which the order numbers 1, 2, 3 and 4, respectively, stand for the quarks 
(antiquarks) $c$, $u$, $\bar{c}$ and $\bar{d}$. It can be drawn a conclusion that the charged 
states $[cu][\bar{c}\bar{d}]$ in the color flux-tube model must not form planar structures but
three-dimensional spatial configurations. The situations of other charged states are similar 
to the states $[cu][\bar{c}\bar{d}]$ and also have three-dimensional spatial configurations. 
The reason for three-dimensional configuration comes from the dynamics of the systems: the color 
flux tube shrinks the distance between any two connected particles to as short a distance as 
possible to minimize the confinement potential energy, while the kinetic motion expands the
distance between any two quarks to as long a distance as possible to minimize the kinetic energy: 
the three-dimension spatial configurations meet this requirement better than a planar one does. 
The four-body confinement potential in the color flux-tube model therefore plays an important role 
in the formation of the three-dimension compact tetraquark states $[Qq][\bar{Q}'\bar{q}']$. 
LQCD calculations on tetraquark states also show that a three-dimensional tetrahedral structure
is favored because a three-dimension configuration is more stable than a planar one~\cite{latt2}. 

Due to the high energies of the tetraquark states, they should eventually decay into several color 
singlet mesons. In the course of the decay, the breakdown of the color flux-tube structures should 
happen first, which leads to the collapses of the three-dimension structures, and then the particles 
adjust the spatial configurations to form decay products by means of the recombination of color flux 
tubes. The decay widthes of the charged states $[Qq][\bar{Q}'\bar{q}']$ are determined by the speeds of
the breakdown and recombination of color flux tubes. The studies of the decay widthes of the charged 
states are in proceeding. This decay mechanism is similar to compound nucleus formation and therefore 
should induce a resonance, which is named as a ``color confined, multi-quark resonance" state in the 
color flux-tube model~\cite{resonance}. It is different from all of those microscopic resonances 
discussed by S. Weinberg~\cite{weinberg}.

\section{summary}
The lowest charged tetraquark states $[Qq][\bar{Q}'\bar{q}']~(Q=c,b,q=u,d,s)$ are studied using the 
variational method GEM in the color flux-tube model with a four-body confinement potential instead
of the sum of the additive two-body confinement. The numerical results indicate that some compact 
resonance states can be formed, in which the four-body confinement potential is an crucial dynamical 
mechanism, and they have three-dimension spatial structures and can not decay into two color singlet 
mesons $Q\bar{q}'$ and $\bar{Q}'q$ but into $Q\bar{Q}'$ and $q\bar{q}'$ by means of the breakdown 
and recombination of the flux tubes. Their decay mechanism is similar to compound nucleus formation 
and therefore should induce a so-called ``color confined, multi-quark resonance" state in the color 
flux-tube model. The newly observed charged states $Z_c(3900)$ and $Z_c(4025)/Z_c(4020)$ can be accommodated 
in the color flux-tube model and can be interpreted as the S-wave tetraquark states $[cu][{\bar{c}\bar{d}}]$ 
with quantum numbers $I=1$ and $J=1$ and 2, respectively. Some predicted charged states are worth being 
searched for in experiments in the future, the studies of the decays of the charged states are in proceeding.

\acknowledgments
{This research is partly supported by the National Science Foundation of China under contracts 
Nos. 11305274, 11047140, 11175088, 11035006, 11265017, 11205091 and by the Chongqing Natural 
Science Foundation under Project No. cstc2013jcyjA00014.}

\end{document}